\documentclass[12pt]{article}
\usepackage{graphicx}

\usepackage{caption,subcaption}

\newcommand\pubnumber{CDF note 11237}
\newcommand\pubdate{\today}

\def\institute{ Institute of Physics of the Czech Academy of Sciences,\\
Prague, Czech Republic}

\def\Title#1{\begin{center} {\Large #1 } \end{center}}
\def\Author#1{\begin{center}{ \sc #1} \end{center}}
\def\Address#1{\begin{center}{ \it #1} \end{center}}

\newcommand\pubblock{\rightline{\begin{tabular}{l} \pubnumber\\
         \pubdate  \end{tabular}}}
\newenvironment{Abstract}{\begin{quotation}  }{\end{quotation}}
\newenvironment{Presented}{\begin{quotation} \begin{center} 
             PRESENTED AT\end{center}\bigskip 
      \begin{center}\begin{large}}{\end{large}\end{center} \end{quotation}}

\def\beq{\begin{equation}}
\def\eeq#1{\label{#1}\end{equation}}
\def\eeqn{\end{equation}}

\def\beqa{\begin{eqnarray}}
\def\eeqa#1{\label{#1}\end{eqnarray}}
\def\eeqan{\end{eqnarray}}

\let\bar=\overbar

\def\Dslash{\not{\hbox{\kern-4pt $D$}}}
\def\dslash{\not{\hbox{\kern-2pt $\del$}}}

\def\msb{{\bar{\ssstyle M \kern -1pt S}}}

\begin{document}

\begin{titlepage}
\pubblock

\vfill
\Title{Top quark properties at Tevatron} 
\vfill
\Author{ Roman Lys\'ak \\ on behalf of the CDF and D0 collaborations}
\Address{\institute}
\vfill
\begin{Abstract}
The latest CDF and D0 experiment measurements of the top quark properties
except the top quark mass are presented. The final
combination of the CDF and D0 forward-backward asymmetry measurements is
shown together with the D0 measurements of the inclusive top quark pair
cross-section as well as the top quark polarization.
\end{Abstract}
\vfill
\begin{Presented}
$10^{th}$ International Workshop on Top Quark Physics\\
Braga, Portugal,  September 18--22, 2017
\end{Presented}
\vfill
\end{titlepage}
\def\thefootnote{\fnsymbol{footnote}}
\setcounter{footnote}{0}

\section{Introduction}

The top quark was discovered at the Tevatron (proton--antiproton
collider) by the CDF and D0 collaborations in 
1995~\cite{Abe:1995hr,Abachi:1995iq}. 
Ever since then the top quark properties have been studied in these
experiments. Most of the data were collected in Run 2 during the years
2001--2011 at a centre-of-mass energy of $\sqrt{s} = 1.96$ TeV  and
correspond to the integrated luminosity of about   
$10\ \rm{fb^{-1}}$ per experiment. Within the Standard Model (SM),  
more than 99\% of top quarks decay into a W boson and b quark
$(t \rightarrow W b)$, so the final states are determined by the W boson
decays: leptonic decays $W \rightarrow \ell\nu$ with the lepton being
an electron, muon or tau, or hadronic decays $W \rightarrow qq$ with
the quarks giving rise to two collimated jets of hadrons in the detector.
The number of the expected top quark pair ($t\bar{t}$)
events passing the typical $t\bar{t}$ event selection 
and corresponding to the total integrated luminosity is about $400$
for the dilepton channel ($t\bar{t} \rightarrow \ell\nu b\ \ell\nu b
$), 2000--4000 for lepton+jets channel 
($t\bar{t} \rightarrow \ell\nu b\ qqb$) depending
on the exact criteria, and about 4000 for all-hadronic channel 
($t\bar{t} \rightarrow qqb\ qqb$).   

Many top quark properties measurements have been performed with
a various precision at the Tevatron.  
These include the measurements
testing the production mechanism of top quark (e.g. inclusive
$t\bar{t}$ cross-section, $t\bar{t}$ forward-backward asymmetry,
$t\bar{t}$ spin correlations), the measurements of the intrinsic
properties of the top quark (e.g. mass, polarization, width, charge),
as well as the measurements of the 
top quark properties related to its decay or the decays of its products
(e.g. $|V_{tb}|$, W boson helicity). The single top quark production
was also observed at the Tevatron.
In the following, only the latest CDF and D0 top quark properties
measurements (except the mass, which was covered in the other presentation
at the conference)
performed within about the last year using the full Tevatron data
statistics are presented.  

\section{Tevatron forward-backward asymmetry}
\begin{figure}[!htb]
\centering

\begin{subfigure}[b]{.48\textwidth}
  \includegraphics[width=\textwidth]{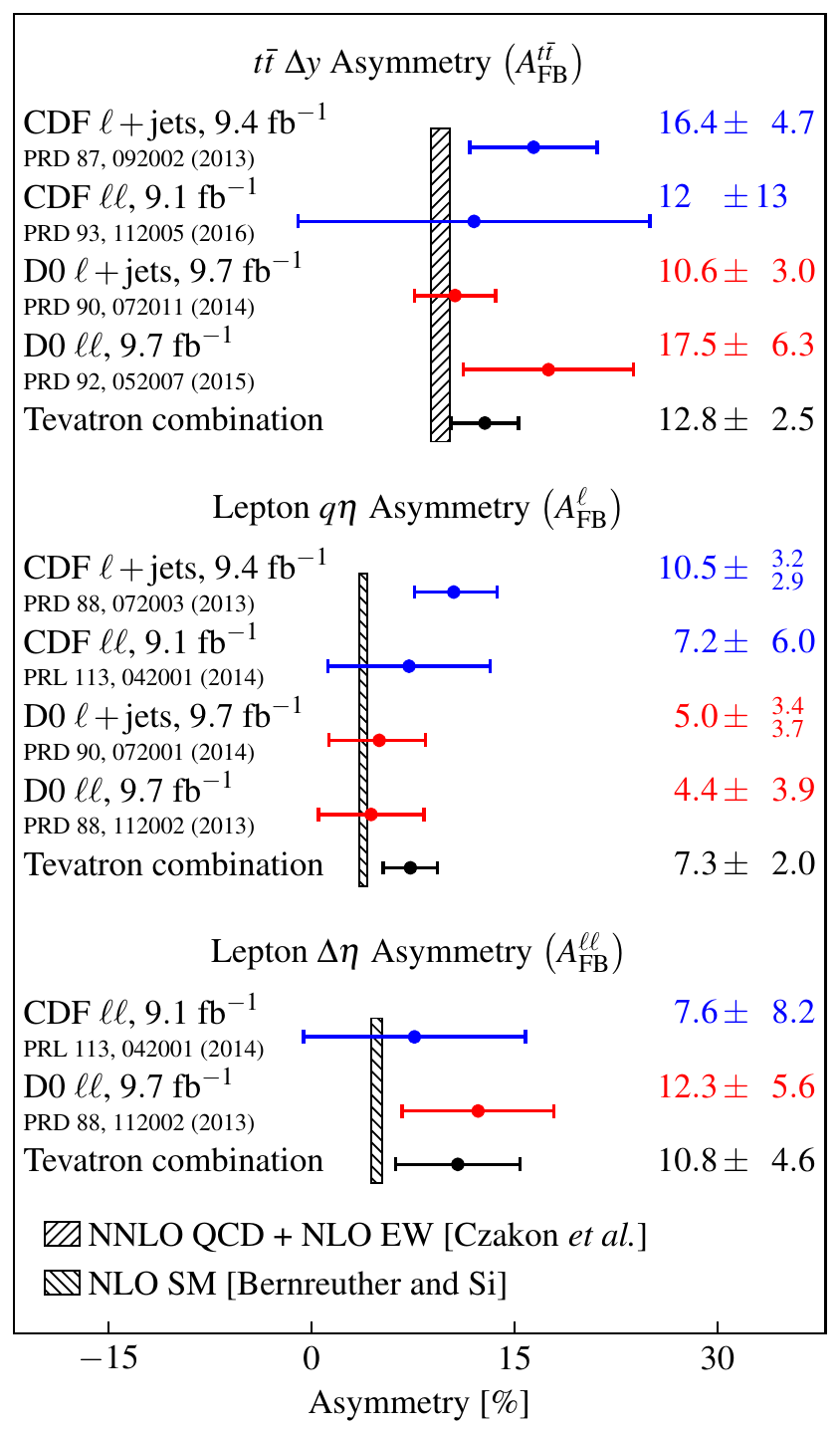}
  \caption{\quad\label{fig:AFB:summary}}
\end{subfigure}
\quad
\begin{subfigure}[b]{.48\textwidth}
  \includegraphics[width=\textwidth]{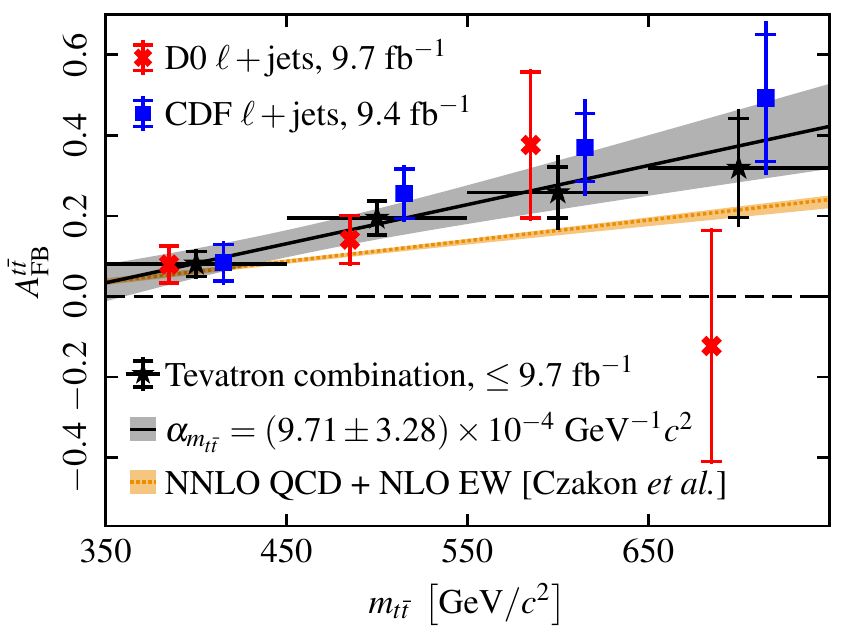}
  \caption{\quad\label{fig:AFB:mtt}}
  \vspace{2ex}
  \includegraphics[width=\textwidth]{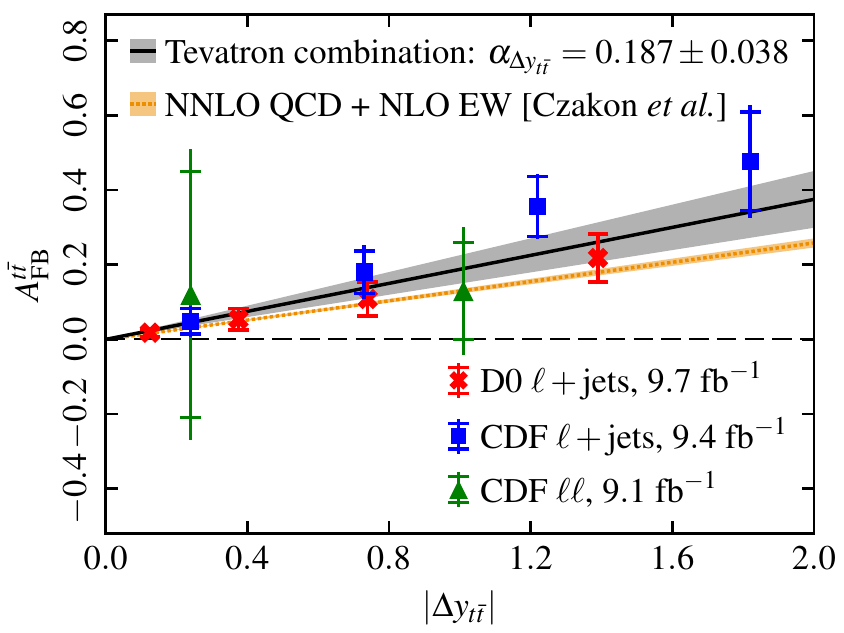}
  \caption{\quad\label{fig:AFB:dy}}
\end{subfigure}
\caption{The summary of CDF and D0 inclusive asymmetry
  measurements is shown in a). The differential measurements
  of $A^{t\bar{t}}_{FB}$ are shown as a function of the invariant mass of
  $t\bar{t}$ pair in b) and $|\Delta y_{t\bar{t}}|$ in c)~\cite{Aaltonen:2017efp}.} 
\end{figure}

\begin{figure}[htb]
\centering
\includegraphics[width=0.46\textwidth]{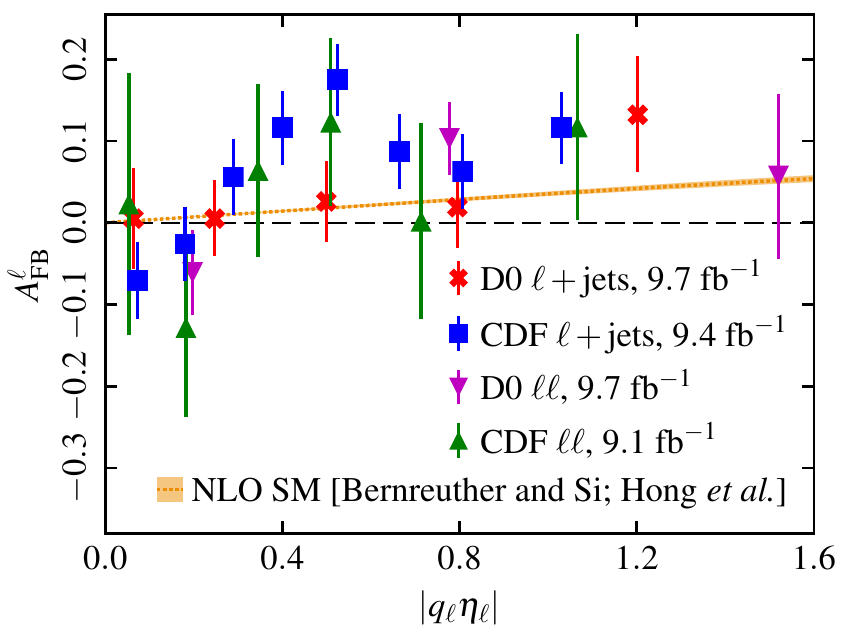}
\includegraphics[width=0.46\textwidth]{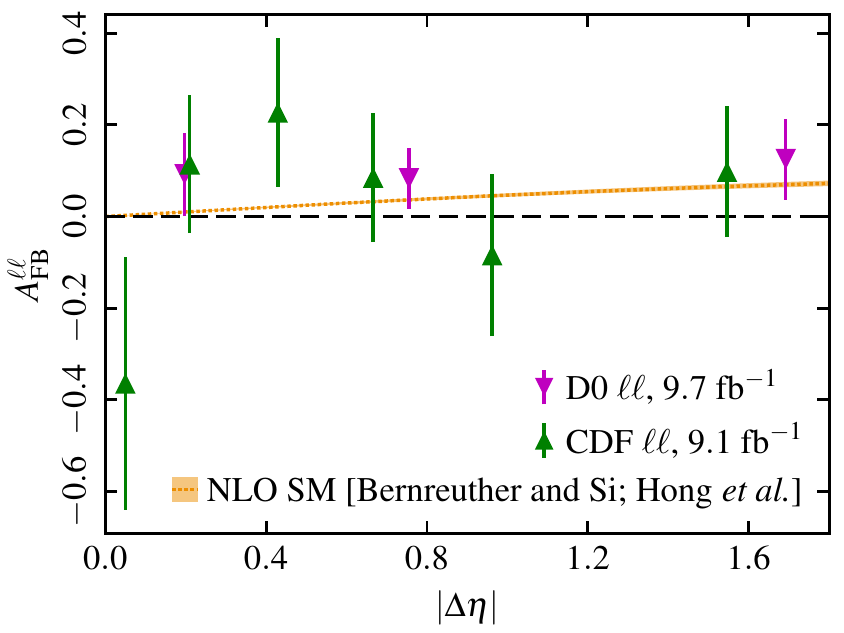}
\caption{The differential $A^{\ell}_{FB}$ measurements as a function of
 $|q_l\eta_l| $ (left) and the differential measurements of
  $A^{\ell\ell}_{FB}$ as a function of $|\Delta \eta|$ (right)~\cite{Aaltonen:2017efp}.
}
\label{fig:AFB:lepton}
\end{figure}

The CDF and D0 experiments performed the combination of their
measurements of the $t\bar{t}$ forward-backward asymmetry from both
dilepton and lepton+jets channels~\cite{Aaltonen:2017efp}.
The forward-backward asymmetry in a given observable $X$ is defined as
\mbox{$A(X) = [N(X>0) - N(X<0)] / [N(X>0) + N(X<0)]$}, where $N(X>0)$
corresponds to number of events with $X>0$.
The asymmetry is defined for three different observables:
$A^{t\bar{t}}_{FB}  = A(X=\Delta y_{t\bar{t}}= y_{top} - y_{antitop})$,
$A^{\ell}_{FB}  = A(X=q_l\eta_l)$ and
$A^{\ell\ell}_{FB}  = A(X=\Delta \eta= \eta_{l^+} - \eta_{l^-})$. 
The combined values for inclusive $A^{t\bar{t}}_{FB}$,
$A^{\ell}_{FB}$, and $A^{\ell\ell}_{FB}$ are summarized together with
the individual input measurements in Fig.~\ref{fig:AFB:summary}.
The differential measurements of $A^{t\bar{t}}_{FB}$ 
as a function of the invariant mass of $t\bar{t}$ as well as $|\Delta
y_{t\bar{t}}|$ are also combined, see Fig.~\ref{fig:AFB:mtt} and Fig.~\ref{fig:AFB:dy}.
The differential $A^{\ell}_{FB}$ and $A^{\ell\ell}_{FB}$ measurements
are just summarized and not combined in Fig.~\ref{fig:AFB:lepton}. 
All results agree with the predictions within 1.3--1.6 standard
deviations ~\cite[and references therein]{Aaltonen:2017efp}. 

\section{D0 $t\bar{t}$ cross-section measurement}

\begin{figure}[htb]
\centering
\includegraphics[width=0.49\textwidth]{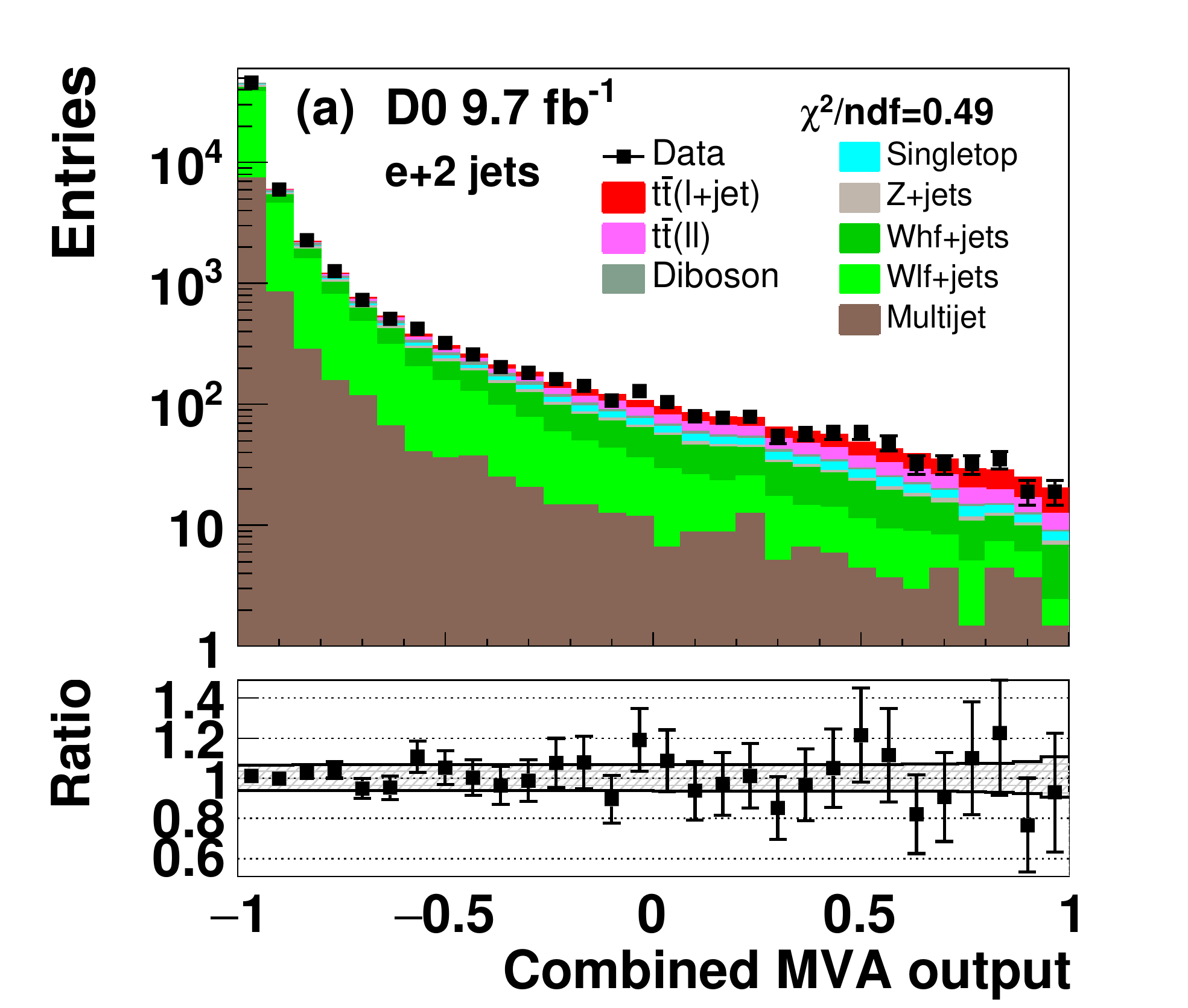}
\includegraphics[width=0.49\textwidth]{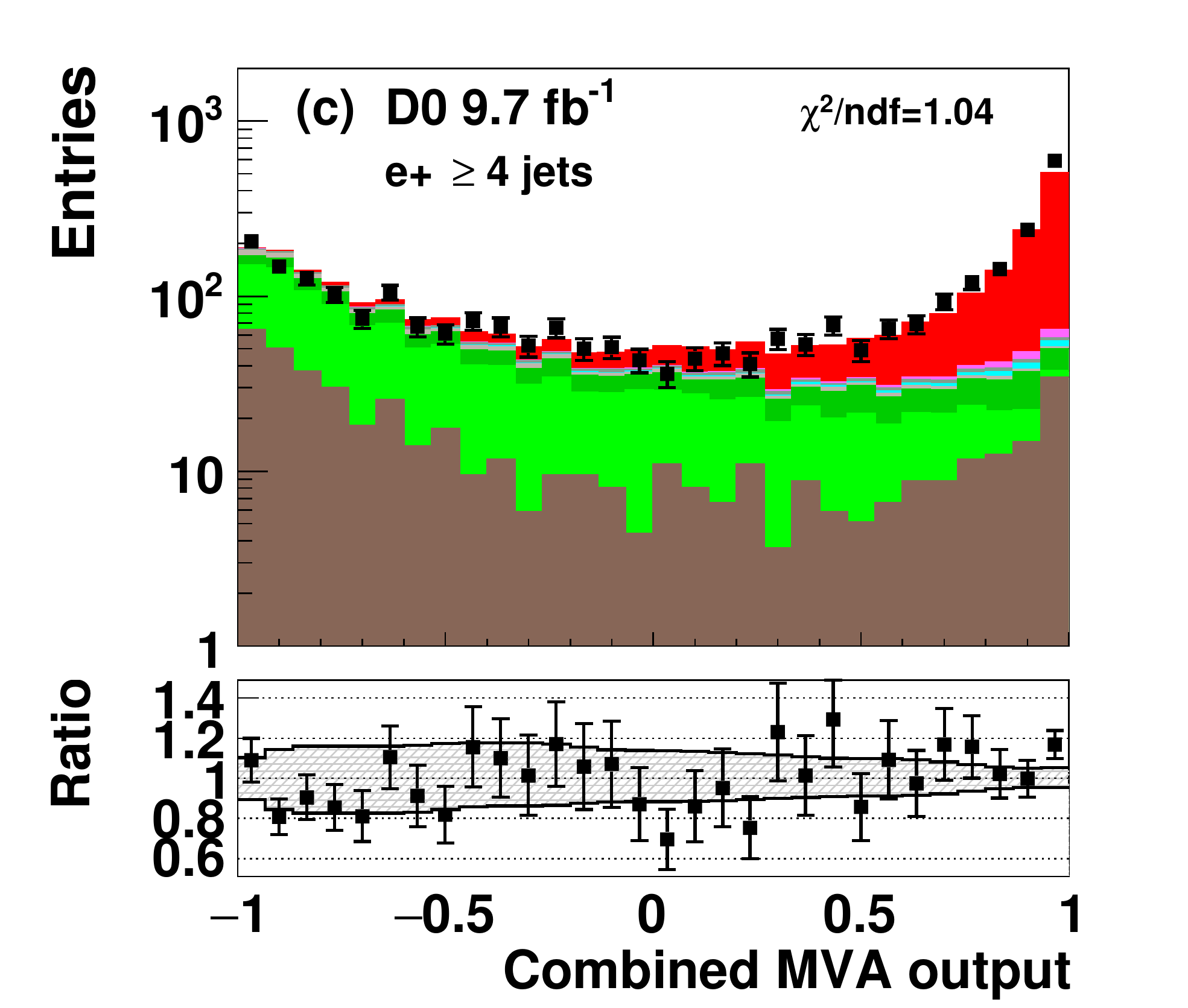}
\caption{The MVA output distributions for channels with electron and two 
  (left) or four jets (right) in the final state~\cite{Abazov:2016ekt}.
}
\label{fig:xsec:mva}
\end{figure}

The D0 experiment measured recently the inclusive $t\bar{t}$
cross-section using the dilepton and 
lepton+jets channels~\cite{Abazov:2016ekt}.
The multivariate (MVA) techniques have been used in both channels to
separate the signal from the background.
The Fig.~\ref{fig:xsec:mva} shows the examples of MVA
output distributions.
The combined result $\sigma_{t\bar{t}} = 7.26 \pm
0.13(stat.) ^{+0.57}_{-0.50}(syst.)~{\rm pb}\ (\delta \sigma / \sigma =
^{+8.1}_{-7.1}\%)$ is compatible with the fully resummed
result at next-to-next-to leading order in perturbative quantum
chromodynamics  
$\sigma_{t\bar{t}} = 7.35^{+0.23}_{-0.27} (\rm{scale} +
\rm{pdf})\ {\rm pb}$~\cite{Baernreuther:2012ws}. The dominant 
sources of systematic uncertainties arrive from the luminosity uncertainty,
the $b$-jet modeling and identification uncertainty, and also from the
signal hadronization modelling uncertainty. 

\section{D0 top quark polarization measurement}
\begin{figure}[htb]
\centering
\includegraphics[width=0.49\textwidth]{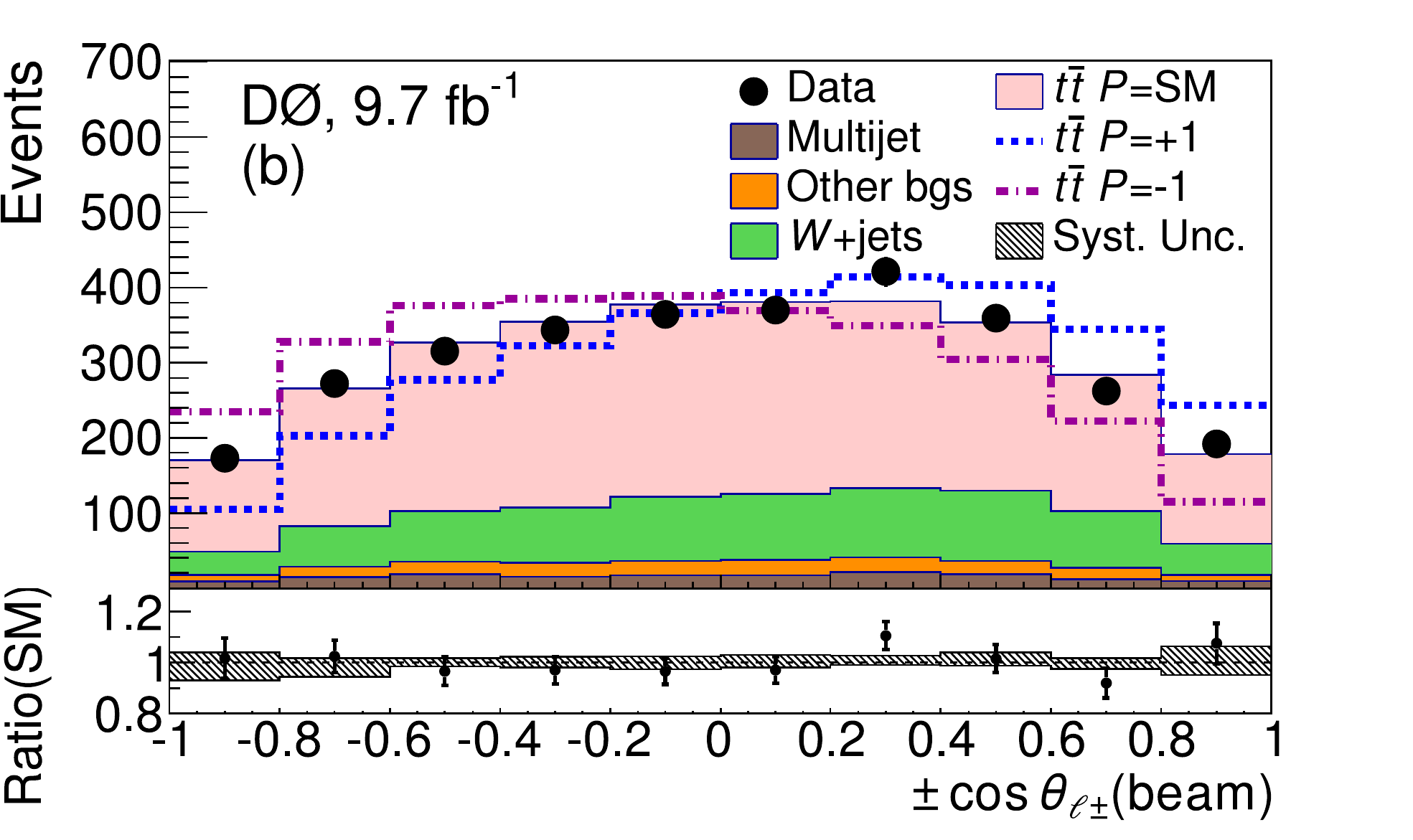}
\includegraphics[width=0.49\textwidth]{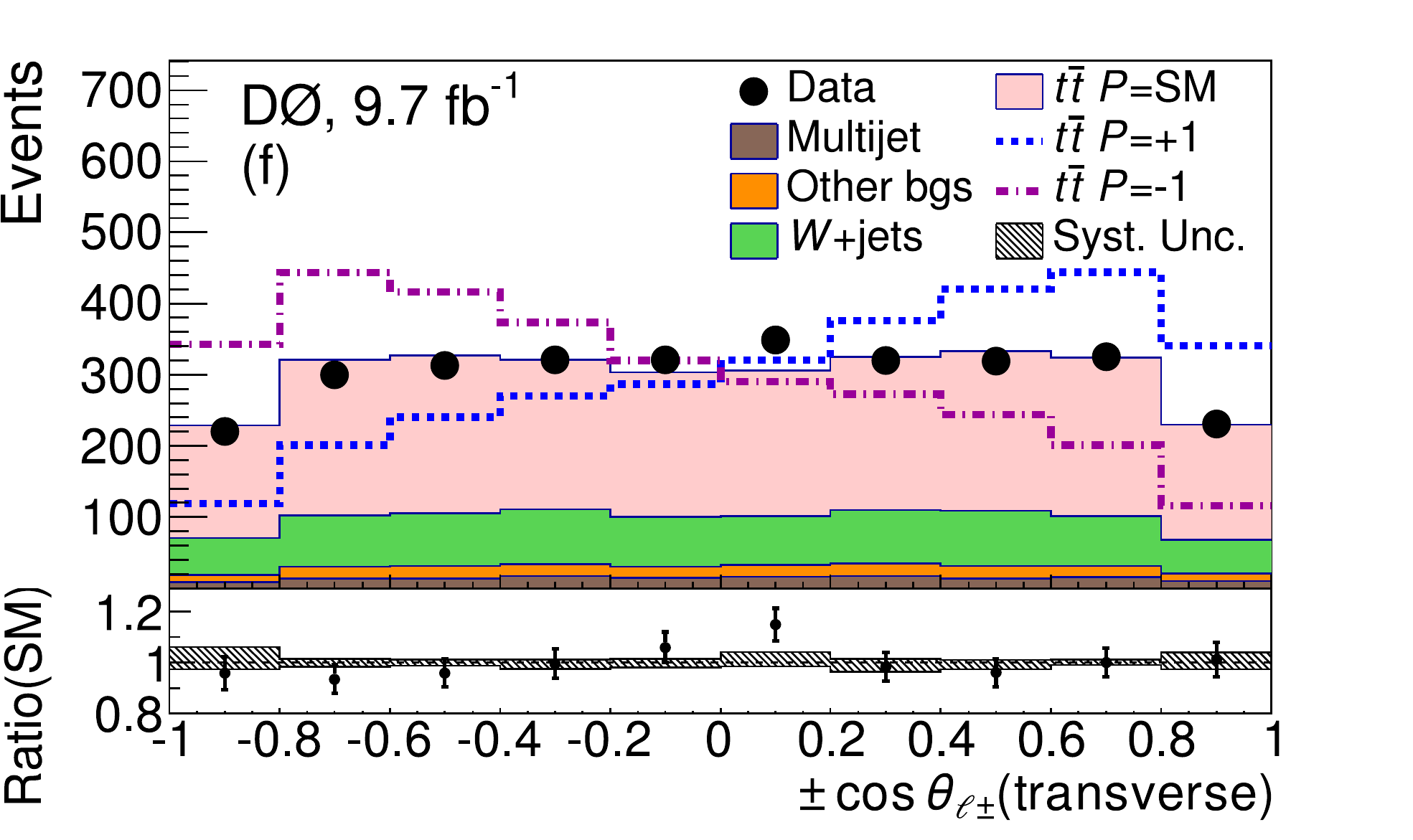}
\caption{The $\cos \theta$ distributions for data, expected
  backgrounds, and signal templates for $P = -1$, SM, and $P = +1$ for
  lepton + 4 jets events. The distributions relative to the
  beam (left) and transverse (right) axis are shown~\cite{Abazov:2016tba}.}
\label{fig:polarisation:cosTheta}
\end{figure}

The D0 experiment performed recently also the measurement of top quark
polarization $P$ in lepton+jets channel~\cite{Abazov:2016tba}.
Within SM, the top quark is produced almost unpolarized in $t\bar{t}$
pair production and this prediction can be tested by measuring the
 angular distributions of the decay products.
In this analysis, the angle $\theta$ between the lepton 
and three different quantization axes is measured. 
 The examples of reconstructed $\cos \theta$ distributions for two
 quantization axes are shown in Fig.~\ref{fig:polarisation:cosTheta}.

\begin{figure}[htb]
\centering
\includegraphics[width=0.53\textwidth]{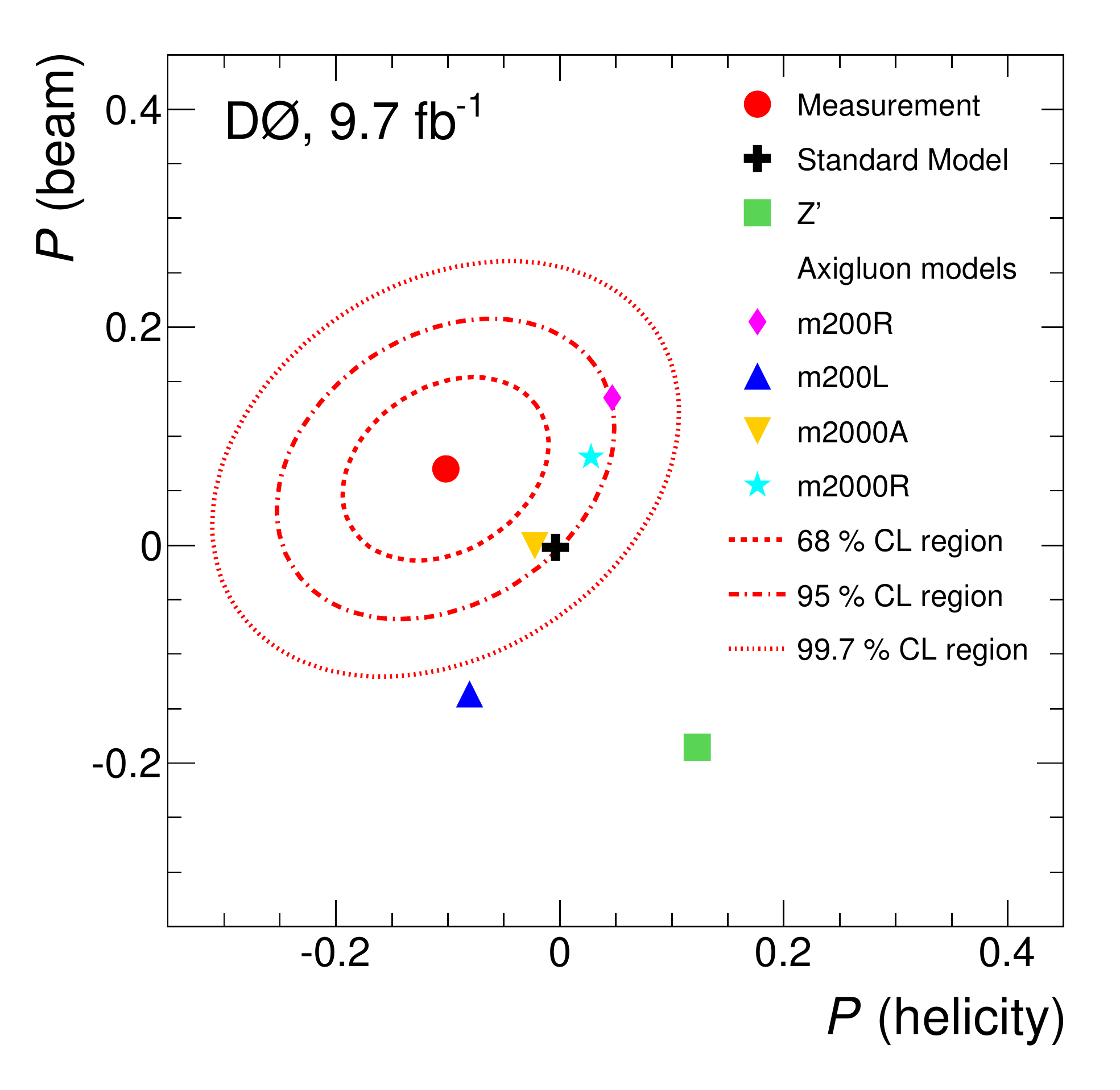}
\caption{Two-dimensional visualization of the top
quark polarizations measured along the beam and helicity axes compared
with the SM and various beyond SM models~\cite{Abazov:2016tba}.} 
\label{fig:polarisation:2D}
\end{figure}

 The final estimate of top polarization is performed by a template fit
 of $\cos \theta$ distribution from data to the combination of
 signal templates with the top quark polarizations $P = \pm 1$ and the
 background  templates. 
 The measurement along the beam axis has been combined with previous
 measurement in the dilepton channel~\cite{Abazov:2015fna}. 
 The results together with the SM predictions are summarized in
 Tab.~\ref{tab:polarization}. All the results are 
consistent with the predictions within 1--2 standard deviations. The
results in beam and helicity axis, which are highly 
correlated, are summarized together also in the
Fig.~\ref{fig:polarisation:2D} where the predictions for various
beyond SM models are also shown.  

\begin{table}[t]
\begin{center}
\begin{tabular}{l|cc}  
Axis &  Measured polarization &  SM prediction\\ \hline
Beam                  & $+0.070 \pm 0.055$   &  $-0.002$\\
Beam--D0 combination   & $+0.081 \pm 0.048$   &  $-0.002$\\
Helicity              & $-0.102 \pm 0.061$   &  $-0.004$\\
Transverse            & $+0.040 \pm 0.035$   &  $+0.011$\\ \hline
\end{tabular}
\caption{The top quark polarization measurements for various
  quantization axes together with the SM predictions~\cite{Abazov:2016tba}.}
\label{tab:polarization}
\end{center}
\end{table}

\section{Summary}
The CDF and D0 experiments are finalizing more than 20 years
of top quark studies using Tevatron data. The final
combination of forward-backward asymmetry measurements have been
already performed while some other combinations, such as CDF and
Tevatron $t\bar{t}$ cross-section, and CDF and Tevatron
top quark mass combinations, are still to be performed.
The precision in most of the top quark measurements at the LHC have
already surpassed that of Tevatron measurements, but some of
them are still competitive, e.g. top quark mass measurements. However,
since LHC is proton--proton collider, a few of the Tevatron
measurements, such as that of inclusive $t\bar{t}$ cross-section and
 $t\bar{t}$ forward-backward asymmetry are unique and represent the legacy measurements.

\end{document}